\documentclass[onecolumn,authoryear]{els-mrw} 

\usepackage{amsmath, amssymb, amsfonts, amsthm, makeidx, graphicx}
\usepackage{txfonts}
\usepackage{helvet}

\definecolor{grey}{rgb}{0.4,0.4,0.4}
\definecolor{dullmagenta}{rgb}{0.4,0,0.4}
\definecolor{darkblue}{rgb}{0,0,0.4}
\definecolor{midblue}{rgb}{0,0,0.7}
\definecolor{midred}{rgb}{0.5,0,0}
\definecolor{orange}{rgb}{1,0.5,0}
\definecolor{lightbrown}{rgb}{0.75,0.5,0.25}
\definecolor{tan}{cmyk}{0.14,0.42,0.56,0}
\definecolor{djunglegreen}{cmyk}{0.99,0,0.52,0}
\definecolor{lightgreen}{rgb}{0,1,0}
\definecolor{olivegreen}{cmyk}{0.64,0,0.95,0.40}
\definecolor{midgreen}{rgb}{0.0,0.675,0.0}
\definecolor{darkgreen}{rgb}{0,0.5,0}
\definecolor{pink}{rgb}{1,0.078,0.57}

\newcommand{\vs}{\vspace}

\renewcommand{\.}{\hspace{0.5mm}}

\newcommand{\la}{\ensuremath{\leftarrow}}

\newcommand{\Brm}{\ensuremath{\mathrm{B}}}

\newcommand{\Drm}{\ensuremath{\mathrm{D}}}

\newcommand{\Hrm}{\ensuremath{\mathrm{H}}}

\newcommand{\Rrm}{\ensuremath{\mathrm{R}}}
\newcommand{\Srm}{\ensuremath{\mathrm{S}}}

\newcommand{\crm}{\ensuremath{\mathrm{c}}}
\newcommand{\drm}{\ensuremath{\mathrm{d}}}

\newcommand{\grm}{\ensuremath{\mathrm{g}}}

\newcommand{\srm}{\ensuremath{\mathrm{s}}}

\newcommand{\MeV}{\ensuremath{\mathrm{MeV}}}

\newcommand{\be}{\begin{equation}}
\newcommand{\ee}{\end{equation}}
\newcommand{\ba}{\begin{eqnarray}}
\newcommand{\ea}{\end{eqnarray}}

\def\ga{\mathrel{\raise.3ex\hbox{$>$\kern-.75em\lower1ex\hbox{$\sim$}}}}
\def\la{\mathrel{\raise.3ex\hbox{$<$\kern-.75em\lower1ex\hbox{$\sim$}}}}

\def\Msun{M_{\odot}}

\def\fPBH{f_{\rm PBH}}

\def\be{\begin{equation}}
\def\ee{\end{equation}}
\def\bea{\begin{eqnarray}}
\def\eea{\end{eqnarray}}

\begin{document}

\chapter{Primordial Black Holes}\label{chap1}

\author[1]{Bernard Carr}%
\author[2]{Florian K{\"u}hnel}%

\address[1]{\orgname{Queen Mary University of London}, \orgdiv{School of Physics \& Astronomy}, \orgaddress{Mile End Road, London E1 4NS, UK}}
\address[2]{\orgname{Max Planck Institute for Physics}, \orgdiv{Physics}, \orgaddress{ Boltzmannstr. 8, 85748 Garching, Germany}}

\articletag{Chapter Article tagline: update of previous edition,, reprint..}

\maketitle


\section*{Key Points}
\label{Key-Points}

\begin{itemize}

	\item Primordial black holes may have formed in the early Universe due to the compression of the big bang, their most likely source being inflationary density fluctuations or some form of cosmological phase transition.\vs{0.5mm}

	\item Numerous constraints from astrophysics and particle physics imply that the fraction of the early Universe collapsing must be tiny but the implied fine-tuning problem can be resolved in some scenarios.\vs{0.5mm}

	\item Such black holes would not grow much and could be small enough for quantum radiation to be important, those below $10^{15}\.$g having evaporated with important consequences for the early Universe.\vs{0.5mm}

	\item Those larger than $10^{15}\.$g could provide the dark matter, explain some of the recently observed gravitational wave events and act as seeds for supermassive black holes and early cosmic structures.\vs{0.5mm}

	\item An attractive scenario which combines all of these features suggests that the black holes have a solar mass and formed at the QCD epoch, although it is also possible that the dark matter comprises black holes of planetary or asteroidal mass.\vs{0.5mm}

	\item Studies of primordial black holes potentially probe four areas of physics: the early Universe, gravitational collapse, high-energy physics and quantum gravity.

\end{itemize}

\begin{glossary}[Nomenclature]
\begin{tabular}{@{}lp{34pc}@{}}
	AGN
		& {\bf A}ctive {\bf G}alactic {\bf N}uclei\\
	BBN
		& {\bf B}ig {\bf B}ang {\bf N}ucleosynthesis\\
	CDM
		& {\bf C}old {\bf D}ark {\bf M}atter\\
	CMB
		& {\bf C}osmic {\bf M}icrowave {\bf B}ackground\\
	EGRET
		& {\bf E}nergetic {\bf G}amma {\bf R}ay {\bf E}xperiment {\bf T}elescope\\
	EROS
		& {\bf E}xp{\'e}rience pour la {\bf R}echerche d'{\bf O}bjets {\bf S}ombres\\
	GW
		& {\bf G}ravitational {\bf W}ave\\
	HSC
		& {\bf H}yper {\bf S}uprime-{\bf C}amera\\
	IMBH
		& {\bf I}ntermediate {\bf M}ass {\bf B}lack {\bf H}ole\\
	IPTA
		& {\bf I}nternational {\bf P}ulsar {\bf T}iming {\bf A}rray\\
	JWST
		& {\bf J}ames {\bf W}ebb {\bf S}pace {\bf T}elescope\\
	KAGRA
		& {\bf Ka}mioka {\bf Gra}vitational-Wave Detector\\
	LIGO
		& {\bf L}aser {\bf I}nterferometer {\bf G}ravitational-Wave {\bf O}bservatory\\
	LVK
		& {\bf L}IGO-{\bf V}irgo {\bf K}ARGRA\\
	LSST
		& {\bf L}arge {\bf S}ynoptic {\bf S}urvey {\bf T}elescope\\
	MACHO
		& {\bf Ma}ssive {\bf C}ompact {\bf H}alo {\bf O}bject\\
	OGLE
		& {\bf O}ptical {\bf G}ravitational {\bf L}ensing {\bf E}xperiment\\
	PBH
		& {\bf P}rimordial {\bf B}lack {\bf H}ole\\
	QCD
		& {\bf Q}uantum {\bf C}hromo {\bf D}ynamics\\
	SMBH
		& {\bf S}uper{\bf M}assive {\bf B}lack {\bf H}ole\\
	WIMP
		& {\bf W}eakly {\bf I}nteracting {\bf M}assive {\bf P}article\\
	UFDG
		& {\bf U}ltra-{\bf F}aint {\bf D}warf {\bf G}alaxy\\	
\end{tabular}
\end{glossary}

\begin{abstract}[Abstract]
In this chapter we first describe the early history of primordial black hole (PBH) research. We then discuss their possible formation mechanisms, including critical collapse from inflationary fluctuations and various types of phase transition. We next describe the numerous constraints on the number density of PBHs from various quantum and astrophysical processes. This was the main focus of research until recently but there is currently a shift of emphasis to the search for evidence for PBHs. We end by discussing this evidence, with particular emphasis on their role as dark matter candidates, sources of gravitational waves and seeds for supermassive black holes and early cosmic structures.
\end{abstract}

\textbf{Keywords}: Black holes; Compact objects; Cosmology; Dark matter; Early universe; Gravitational lensing; Gravitational wave astronomy; Hawking radiation; Primordial black holes

\section*{Introduction}
\label{sec:Introduction}

\noindent According to general relativity, when a region of mass $M$ is compressed within its Schwarzschild radius, $R_{\Srm} \equiv 2\.GM / c^{2}$, it forms a black hole. While these objects could theoretically exist across many mass scales, natural astrophysical processes can only create them above a solar mass. Black holes several times heavier than the Sun can form at the endpoint of stellar evolution, with our Galaxy's disc alone containing hundreds of millions of them. Stars larger than a few hundred solar masses, above the pair-instability mass gap, can potentially form Intermediate Mass Black Holes (IMBHs) through direct collapse. This is the category that might include the Universe's earliest stars. Supermassive Black Holes (SMBHs), with masses from $10^{6}\.\Msun$ to $10^{10}\.\Msun$, reside in galactic nuclei and would power quasars. The one at the center of the Milky Way has a mass of $4 \times 10^{6}\.\Msun$.

Another type of black hole could have formed during the Universe's earliest moments and these are called ``primordial''. Comparing the cosmological density at a time $t$ after the Big Bang with the density needed for matter to collapse within its Schwarzschild radius, implies that primordial black holes (PBHs) would have of order the cosmic horizon mass, $M \sim c^{3}\.t/G$, at formation. Their possible masses therefore span a huge range: from the Planck mass ($M_{\rm Pl} \sim 10^{-5}\.\grm$) if created at the Planck time ($10^{-43}\.$s), to a solar mass ($1\.\Msun$) if formed during the quantum chromodynamics (QCD) epoch ($10^{-5}\.$s) and $10^{5}\.\Msun$ if generated at $t \sim 1\.$s. So PBHs could exist below the traditional solar-mass threshold. They could also be small enough for Hawking radiation to be important, those lighter than Earth being hotter than the cosmic microwave background (CMB) and those below $10^{15}\.$g evaporating within the current age of the Universe.

\begin{figure}[t]
	\centering
	\includegraphics[width=0.70\textwidth]{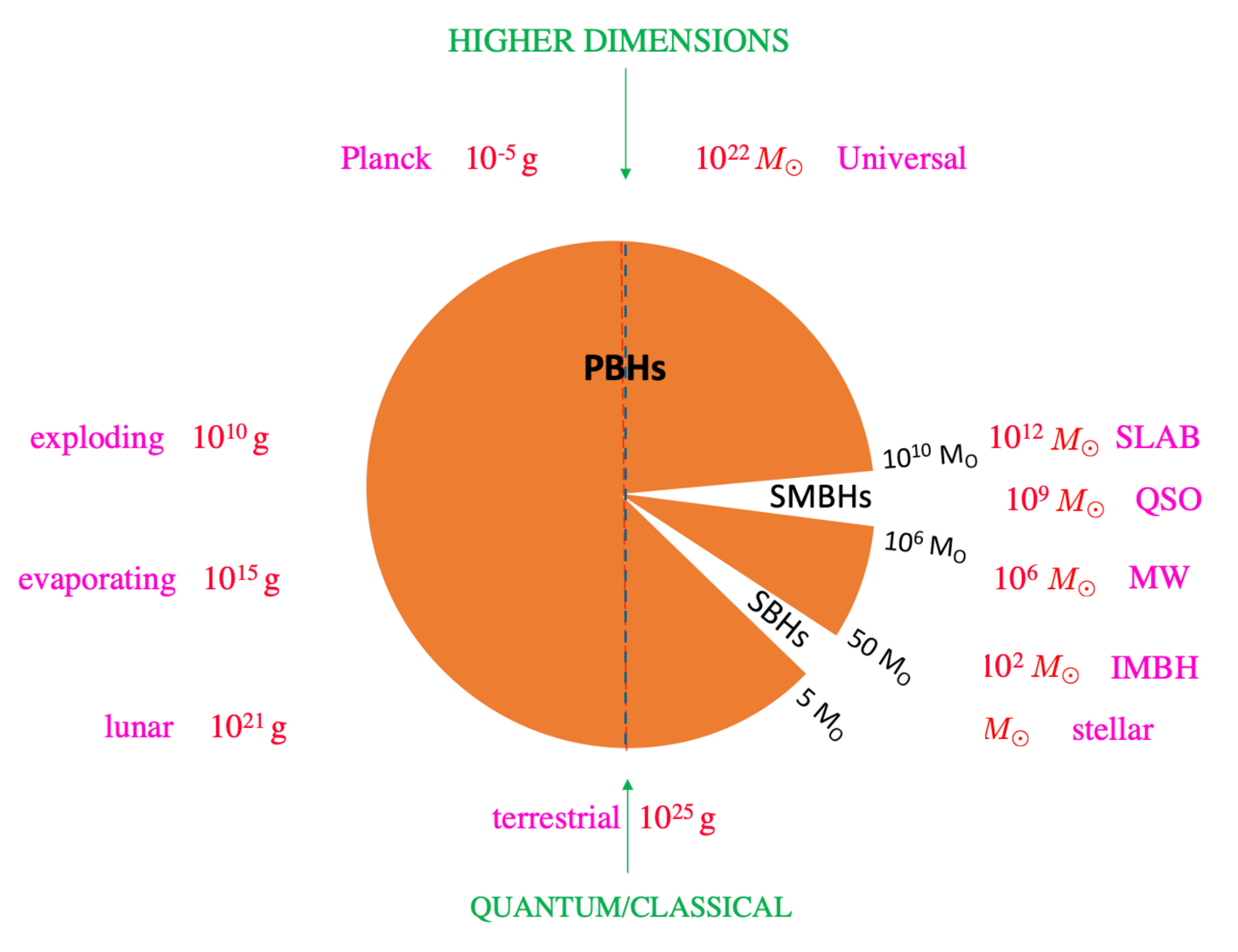}
	\caption{%
		This indicates the mass and size of 
		various types of black holes, with the 
		division between the micro and macro 
		domains being indicated by the vertical 
		line. QSO stands for Quasi-Stellar 
		Object, MW for Milky Way, IMBH for 
		Intermediate Mass Black Hole, SLAB for 
		Stupendously Large Black Hole. Stellar 
		black holes (SBHs) and supermassive black 
		holes (SMBHs) occupy only the small 
		slivers shown in white, whereas PBHs 
		occupy the much wider range shown in 
		orange.}
\label{fig:urob}
\end{figure}

The wide mass range of black holes and their crucial role in linking macrophysics and microphysics is summarised in Figure \ref{fig:urob}. The edge of the orange circle can be regarded as a sort of ``clock'' in which the scale changes by a factor of $10$ for each minute, from the Planck scale at the top left to the scale of the observable Universe at the top right. The top itself corresponds to the Big Bang because at the cosmological horizon distance one is peering back to an epoch when the Universe was very small, so the very large meets the very small there. The various types of black holes are labelled by their mass and positioned according to their Schwarzschild radius. On the right are the astrophysical black holes, with the well-established stellar and supermassive ones corresponding to the segments between $5$ and $50\.\Msun$ and between $10^{6}$ and $10^{10}\.\Msun$, respectively. On the left{\,---\,}and possibly extending to the right{\,---\,}are the more speculative PBHs. 

The vertical line between the bottom (planetary-mass black holes) and the top (Planck-mass black holes) provides a convenient division between the microphysical and macrophysical domains. Quantum emission is suppressed by accretion of the CMB to the right of the bottom point, so this might be regarded as the transition to classical black holes. The effects of extra dimensions could be important at the top, especially if they are compactified on a scale much larger than the Planck length. In this context, there is a sense in which the whole Universe might be a PBH; this is because in brane cosmology (in which one extra dimension is extended) the Universe can be regarded as emerging from a five-dimensional black hole. Although we cannot be certain that PBHs formed at all, research into their consequences potentially probes four distinct areas of physics:
	(1) the early Universe ($M < 10^{15}\.$g);
	(2) gravitational collapse ($M > 10^{15}\.$g);
	(3) high-energy physics ($M \sim 10^{15}\.$g);
	(4) quantum gravity ($M \sim 10^{-5}\.$g).

\section*{Early Work on PBH Formation, Accretion and Evaporation}
\label{sec:Early-Work-on-PBH-Formation,-Accretion-and-Evaporation}

\noindent The possibility of PBH formation emerges from two key density expressions: the cosmic density following the Big Bang, $\rho \sim 1 / ( G t^{2} )$, and the critical density needed for an object to collapse within its Schwarzschild radius, $\rho \sim c^{6} / ( G^{3} M^{2} )$. This implies an initial PBH mass:
\begin{equation}
	\label{eq:Moft}
	M
		\sim
			\frac{ c^{3}\mspace{1.5mu}t }{ G }
		\sim
			10^{15}\mspace{-1mu}
			\left(
				\frac{ t }{ 10^{-23}\.\srm }
			\right)
			\.\grm
			\, ,
\end{equation}
this being of order the mass contained within the cosmic horizon at time $t$. This relation was used in a seminal paper by~\citet{Hawking:1971ei}, in which he proposed that Planck-mass PBHs would carry electric charge, enabling them to capture electrons or protons to create atom-like structures. He suggested these ``atoms'' might be detectable through tracks in bubble chambers and that they might also be concentrated in stellar cores. Later, it was realised that such small black holes would lose their charge through quantum effects but the notion that PBHs might be swallowed by stars has recently become topical.

Interestingly, an earlier analysis by~\citet{1967SvA....10..602Z} had led them to conclude that such objects could not exist because they would undergo runaway accretion. Their skepticism stemmed from a Bondi accretion calculation which suggested that PBHs formed with the horizon mass (as expected) would grow at the same rate as the cosmic horizon, reaching $10^{17}\.\Msun$ by the end of the radiation-dominated era. Since the existence of such huge black holes is precluded, this suggested that PBHs never formed. However, this argument neglected the cosmic expansion, which is important for PBHs with the horizon size and would inhibit accretion. Later~\citet{Carr:1974nx} showed that there is no solution in general relativity in which a black hole formed from local collapse can grow as fast as the cosmic horizon. Furthermore, the black hole would soon become much smaller than the horizon, at which point the Bondi formula {\it should} apply and implies that it should not grow much at all. 

The removal of the concerns raised by Zeldovich--Novikov reinvigorated PBH research and led to 
Hawking's revolutionary discovery that black holes emit thermal radiation with a temperature:
\begin{equation}
	T_{\rm BH}
		=
			\frac{ \hbar\mspace{1.5mu}c^{3} }
			{ 8\pi\mspace{1.5mu}G M\mspace{1.5mu}k_{\Brm} }
		\sim
			10^{-7}
			\left(
				\frac{ M }{ M_{\odot} }
			\right)^{-1}
			{\rm K}
			\, .
			\label{eq:temp}
\end{equation}
This results in their complete evaporation over a time:
\begin{equation}
 \tau( M )
 	\approx
			5120\.\pi\,
			\frac{ G^{2}\mspace{1mu}M^{3} }
			{ \hbar\mspace{1.5mu}c^{4} }
		\sim
			10^{67}
			\left(
				\frac{ M }{ M_{\odot} }
			\right)^{3}
			{\rm yr}
			\, .
\end{equation}
PBHs of initial mass $M_{*} \sim 10^{15}\.$g, formed at $10^{-23}\.$s with the size of a proton, would be completing their evaporation in the present era, while less massive ones would have vanished earlier. However, PBHs heavier than Earth ($10^{27}\.$g) would be cooler than the CMB, causing them to grow through accretion rather than shrink through evaporation. Hawking's discovery has not yet been confirmed experimentally and there remain major conceptual puzzles associated with the process. Nevertheless, it is generally recognised as one of the key developments in 20th-century physics because it beautifully unifies general relativity, quantum mechanics and thermodynamics. The fact that Hawking was only led to this discovery through contemplating the properties of PBHs illustrates that it has been useful to study them even if they never actually formed.

This period also saw the emergence of more detailed models of PBH formation. The high density of the early Universe does not guarantee PBH formation because one also needs density fluctuations, so that overdense regions can eventually stop expanding with the background. However, only regions larger than the Jeans length at maximum expansion could collapse against the pressure. This is approximately $\sqrt{w}$ times the horizon scale for an equation of state is $p = w\.\rho$. For PBHs forming from Gaussian density fluctuations with root-mean-square amplitude $\epsilon( M )$ at horizon crossing, this implies that the fraction of the Universe collapsing into PBHs of mass $M$ should be~\citep{Carr:1975qj}
\begin{equation}
	\beta( M )
		\sim
			\epsilon( M )\.
			\exp\mspace{-1.5mu}
			\left[
				- \frac{ w^{2} }
				{ 2\.\epsilon( M )^{2} }
			\right]
			\. .
	\label{eq:beta}
\end{equation}
One expects $w = 1/3$ during the radiation-dominated era and it was initially expected that $\epsilon( M )$ should be scale-invariant. In this case, $\beta$ should also be scale-invariant and the PBH mass function should fall as $M^{-5/2}$. However, as discussed later, this does not apply for the fluctuations expected in the inflationary scenario since these are not exactly scale-invariant. 

On cosmological scales the amplitude of the fluctuations at horizon crossing is around $10^{-5}$, so the exponential dependence in Eq.~\eqref{eq:beta} implies that the collapse fraction is tiny unless $\epsilon$ is much enhanced on small scales. Observations also {\it require} this since, if the current density parameter of PBHs which form at redshift $z$ is $\Omega_{\rm PBH}$ in units of the critical density, the initial collapse fraction must have been
\begin{equation}
	\beta 
		=
			\frac{ \Omega_{\rm PBH} }
			{ \Omega_{\Rrm} }\.
			( 1 + z )^{-1}
		\sim
			10^{-6}\.
			\Omega_{\rm PBH}
			\left(
				\frac{ t }{ \srm }
			\right)^{1/2}
		\sim
			10^{-18}\.
			\Omega_{\rm PBH}
			\left(
				\frac{ M }{ 10^{15}\.\grm }
			\right)^{1/2}
			,
\end{equation}
where $\Omega_{\Rrm} \approx 10^{-4}$ is the current density parameter of radiation and we have used Eq.~\eqref{eq:Moft} at the last step. The $( 1 + z )$ factor arises because the radiation density scales as $( 1 + z )^{4}$, whereas the PBH density scales as $( 1 + z )^{3}$. So $\beta$ must be tiny even if PBHs provide all of the dark matter. This is a potential criticism of the PBH dark matter proposal, since it requires fine-tuning of $\beta$ and even greater fine-tuning of $\epsilon$. There is also the puzzling feature that the PBH and baryon densities are very close if PBHs provide the dark matter, although (as discussed later) there is one scenario in which this arises naturally.

The constraints on $\beta( M )$ are particularly strong on the scales for which evaporation is important. Since PBHs with a mass of $10^{15}\.$g would be producing photons with energy of order $100\.$MeV at the present epoch, the observational limit on the $\gamma$-ray background intensity at $100\.$MeV immediately implies that their density could not exceed $10^{-8}$ times the critical density~\citep{Page:1976wx}. This means that there is little chance of detecting black hole explosions at the present epoch, which would have confirmed the existence of both PBHs and Hawking radiation. Nevertheless, the evaporation of PBHs smaller than $10^{15}\.$g could still have many interesting cosmological consequences, each associated with a different type of emitted particle, and it also implied interesting constraints on the collapse fraction $\beta( M )$. 

The evaporation constraints were first brought together by~\citet{1979A&A....80..104N}. The strongest one is the $\gamma$-ray limit associated with the $10^{15}\.$g PBHs. Others are associated with the generation of entropy and modifications to the cosmological production of light elements, PBHs with $ M \sim 10^{10}\,\grm$ having a lifetime $ \tau \sim 10^{3}\.\srm$ and therefore evaporating at the big bang nucleosynthesis (BBN) epoch. Injection of high-energy neutrinos and antineutrinos changes the weak interaction freeze-out time and hence the neutron-to-proton ratio at the onset of BBN, modifying ${}^4\mathrm{He}$ production. PBHs with $M = 10^{10}$\,--\,$10^{13}\.\grm$ evaporated after BBN, increasing the baryon-to-entropy ratio present at nucleosynthesis and resulting in overproduction of ${}^4\mathrm{He}$ and underproduction of $\Drm$. Emission of high-energy nucleons and antinucleons increases the primordial deuterium abundance due to the capture of free neutrons by protons and spallation of ${}^4\mathrm{He}$. The emission of photons by PBHs with $M > 10^{10}\.\grm$ dissociates the deuterons produced in nucleosynthesis. The associated constraints on $\beta( M )$ are discussed later.

PBHs were also invoked to explain certain observations. For example, evaporating PBHs of around $10^{15}\.$g might explain the 511 keV annihilation line radiation from the Galactic centre or antiprotons in cosmic rays.
However, such evidence was not strong, so attention soon switched to PBHs more massive than $10^{15}\.$g which are unaffected by Hawking radiation. In particular, it was soon realised that these might be dark-matter candidates~\citep{1975Natur.253..251C} and that sufficiently large ones could generate cosmic structures through their Poisson effect~\citep{Meszaros:1975ef}.

\section*{Formation Scenarios}
\label{sec:Formation-Scenarios}

\noindent The early work assumed that PBHs form from primordial inhomogeneities but did not specify the source of the inhomogeneities. In the following decades researchers explored the possible sources very extensively. They also used numerical collapse calculations, thereby improving the earlier heuristic criterion for PBH formation. The numerical work is largely independent of the source of the fluctuations and we discuss both these topics below.

\paragraph*{PBH Formation from Inflationary Perturbations}
\label{subsec:inflation}
\vs{2mm}

\noindent Inflation describes an epoch of extremely rapid (generally exponential) expansion in the early Universe.
Since this exponentially dilutes any pre-existing PBHs, the horizon mass when inflation ends leads to a minimum mass for detectable ones. Observations of tensor-mode large-scale CMB temperature variations constrain the post-inflation reheating temperature to below $10^{16}\.$GeV, corresponding to a minimum PBH mass of around $\sim 1\,\grm$. However, the quantum behaviour of the inflaton field would also generate density fluctuations. For single-scalar-field slow-roll inflation, the perturbation amplitude is $\epsilon \propto V^{3/2} / V^{\prime}$, where $V$ is the potential energy and ${V}^{\prime}$ its gradient, which raises the question of whether this might create PBHs. 

The tiny density fluctuations at CMB scales and Eq.~\eqref{eq:beta} imply that one cannot produce even a single PBH within our current horizon. Creating a significant number of PBHs therefore requires perturbations that either increase on smaller scales (a blue spectrum) or exhibit a peak at some scale. Both of these are possible with single-field inflation but one needs to go beyond the slow-roll approximation. 
Early studies of PBH formation in models with blue spectra showed that avoiding their overproduction places a tighter constraint on the spectral index of the primordial perturbations than CMB observations, while
models with a plateau in the potential, where $V^{\prime} \rightarrow 0$, could also generate PBHs. However, more complicated models were also proposed, such as
hybrid inflation; this has two fields, with one having large quantum fluctuations as it undergoes a phase transition which ends inflation. 

Researchers later investigated numerous inflationary models capable of generating PBHs without violating the perturbation constraints on cosmological scales. These included hilltop inflation, where the field evolves from a local maximum, double inflation, characterised by two inflationary phases, multifield models, with rapid turns in field space, and models with a period of ultra-slow-roll inflation driven by an inflection point or plateau. Quantum diffusion{\,---\,}inevitable in many such scenarios{\,---\,}leads to a tail for the distribution function much shallower than Gaussian, thereby increasing the likelihood of PBH formation. The reheating period following inflation could also generate significant perturbations.

\paragraph*{PBH Formation from Phase Transitions}
\vs{2mm}

\noindent PBHs could be generated by colliding bubbles during a first-order phase transition, with masses comparable to the horizon mass then. However, this requires precise fine-tuning of bubble nucleation rates to ensure sufficient collisions while preventing the phase transition from completing too rapidly. Cosmic phase transitions can also generate one-dimensional topological structures called cosmic strings These can self-intersect to form oscillating loops which can collapse into PBHs if they contract below their Schwarzschild radius. The resulting PBH mass is close to the loop mass, which in turn relates to the horizon mass. Since the loop collapse rate remains constant over time, this mechanism produces a PBH mass distribution with $\drm n / \drm M \propto M^{-5/2}$, just like the distribution from scale-invariant density perturbations. The fraction of loops collapsing to PBHs is very sensitive to the string tension $(G \mu)$ and there are strong constraints on this parameter from other observations. More exotic mechanisms for the production of PBHs also have been proposed, such as quark confinement,
Q-balls produced by the fragmentation of scalar field condensate,
long-range forces mediated by scalar fields
or QCD colour-charge effects.

Conventional cosmological models assume that radiation dominated the Universe's energy content from its earliest moments until matter-radiation equality ($t_{\rm eq} = 1.7 \times 10^{12}\.\srm$). However, alternative scenarios{\,---\,}involving long-lived non-relativistic particles that eventually decay{\,---\,}would allow an early period of matter dominance~\citep{Khlopov:1980mg}. A matter-dominated phase could also follow after inflation if the Universe reheats gradually. These scenarios modify the requirements for PBH formation and generally make it easier. However, density perturbations must maintain near-perfect spherical symmetry to produce PBHs, thereby avoiding the pancake or cigar configurations typical of asymmetric collapse. Also, the matter must fall within its Schwarzschild radius before central caustics form. Numerical studies of PBH formation during matter-domination find that the initial PBH mass fraction in this case is $\beta \approx 0.21\.\epsilon^{13/2}$ for $\epsilon \ll 1$~\citep{Harada:2016mhb}. Since angular momentum is significant for PBHs formed during matter-domination, they have larger spins than those formed during radiation-domination. Even if the pressure does not go to zero, as in a matter-dominated Universe, the threshold for PBH formation, $\delta_{\crm}$, will decrease whenever there is a reduction in pressure and we discuss a realisation of this later.

\paragraph*{Numerical Calculations and Critical Collapse}
\vs{2mm}

\noindent The first numerical studies of PBH formation were carried out by \citet{1978SvA....22..129N}, modelling overdense regions as $k = +1$ Friedmann models matched to a $k = 0$ background by a vacuum region. These roughly confirmed the simple analytic prediction but included the effect of pressure gradients, resulting in PBHs somewhat smaller than the horizon. Later numerical calculations were influenced by the discovery of the phenomenon of critical collapse. This introduces a specific relationship between a black hole's mass and the perturbation from which it forms: $M \propto M_{\Hrm} ( \delta - \delta_{\crm} )^{\gamma}$, where the scaling exponent $\gamma$ is constant for a given equation of state. \citet{Niemeyer:1997mt} pointed out that this applies to PBH formation and calculated the resulting PBH mass function. They also carried out simulations of PBH formation for differently shaped perturbations. Subsequent work in a series of papers by Musco and collaborators 
has explored critical collapse more carefully, verifying the mass scaling for small ($\delta - \delta_{\crm}$) and studying a range of equation of state parameters ($0 < w < 0.6$). Later investigations showed that the collapse threshold for PBH formation depends on the specific shape of density perturbations
and it is universal when expressed in terms of the average compaction function,
which relates to the Schwarzschild gravitational potential.

\paragraph*{Extended Mass Functions, Clustering and Non-Gaussianity}
\label{subsec:emf}
\vs{2mm}

\noindent The early analyses treated PBHs as uniformly distributed objects of identical mass (i.e.~with a delta-function mass distribution). However, theoretical formation mechanisms generally predict both a spectrum of masses and enhanced spatial clustering at small scales. For example, inflation may produce a lognormal mass function~\citep{PhysRevD.47.4244} and critical collapse produces a power-law low-mass tail. This is a two-edged sword as regards the observational constraints: while limits on the total PBH density can be weakened by spreading out the mass, a large density being permitted at one mass scale may not exclude their contravention on some other scale. A general method for applying constraints for a monochromatic mass function to an extended mass function has been presented.

Statistical clustering of PBHs arising from the Poisson fluctuations has attracted considerable research attention. A particularly intriguing consequence is the accelerated formation of the first baryonic structures compared to conventional cosmological models~\citep{2016ApJ...823L..25K}. Numerical simulations of cluster evolution in scenarios where PBHs constitute a portion of dark matter have been conducted by \citet{Inman:2019wvr}. These clustering effects modify existing PBH abundance constraints in complex ways. For diffuse clusters originating from Gaussian density fluctuations, microlensing constraints remain largely unchanged.
For compact clusters (potentially arising from non-Gaussian perturbations), the microlensing constraints are weakened but other constraints are strengthened. 

Since PBHs form from rare large perturbations, their abundance depends sensitively on the shape of the tail of the probability distribution of the perturbations. For the flat inflaton potential required to generate large perturbations in single-field models, the quantum fluctuations are expected to generate a non-Gaussian probability distribution. However, even when the initial curvature perturbations are Gaussian, the nonlinear relationship connecting density and curvature fluctuations inevitably produces non-Gaussian distributions for large density perturbations.

\paragraph*{Fine-Tuning}
\vs{2mm}

\noindent There are three fine-tuning problems:
	(1) Even if PBHs have significant cosmic abundance (e.g.~providing the dark matter), the collapse fraction $\beta$ must be tiny. 
	(2) If they form from primordial density perturbations, the exponential dependence of $\beta$ on $\epsilon$ implies even stronger tuning of the amplitude of these perturbations. 
	(3) If the perturbations arise from inflation, this in turn requires fine tuning of the underlying inflationary parameters.
There is an elegant solution to the first problem. If the PBHs form during the QCD epoch and provide the dark matter, then the collapse fraction $\beta$ naturally approximates the cosmic baryon-to-photon ratio $\eta$, which suggests that there may be a connection between these two quantities. Indeed, \citet{Garcia-Bellido:2019vlf} have proposed a scenario in which PBH formation at the QCD epoch naturally explains this connection. The energy released by the collapsing matter generates an outgoing shock which is hot enough for baryogenesis to produce a {\it local} baryon asymmetry of order $1$ and this implies $\beta \sim \eta$ after the baryons have diffused throughout the Universe. This also explains why the baryons and dark matter have comparable densities. It does not explain the actual value of $\beta$ but this might have an anthropic explanation since galaxies cannot form if the dark matter fraction is too large or too small.

\section*{Observational Constraints}
\label{sec:Observational-Constraints}

\begin{figure}[t]
	\centering
	\includegraphics[width=0.7\textwidth]{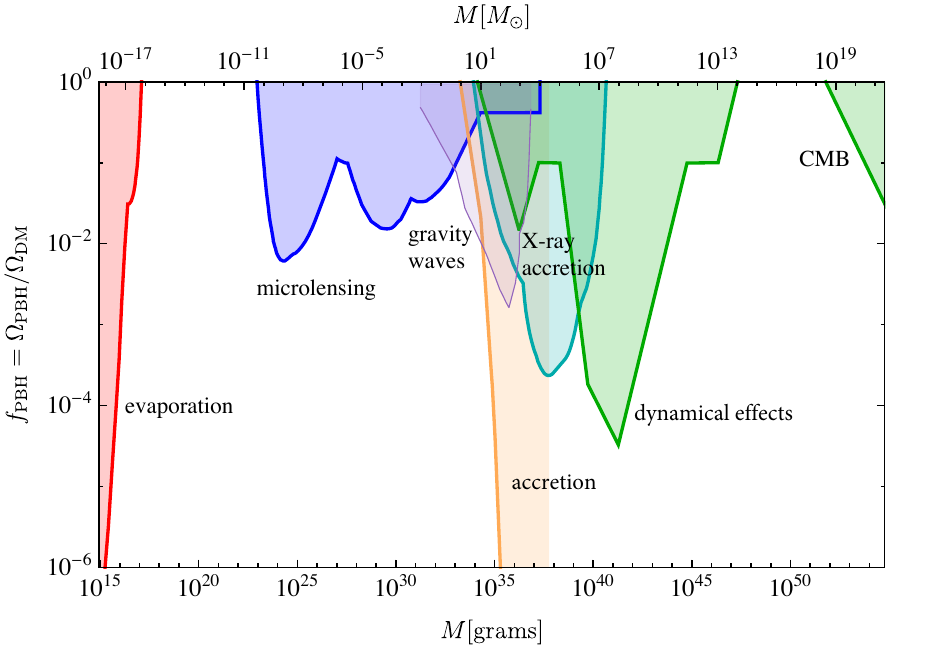}
	\caption{%
		Constraints on the fraction of DM in 
		the form of PBHs, $f_{\rm PBH}$, as a 
		function of mass, $M$, assuming all PBHs 
		have the same mass. The shaded regions 
		are excluded under standard assumptions. 
		The bounds shown are from 
			evaporation (red), 
			microlensing (blue), 
			gravity-wave (purple), 
			accretion (yellow),
			X-ray accretion (light blue),
		and 
			dynamical (green) effects, including 
			the CMB dipole.
				}
	\label{fig:constraints}
\end{figure}

\noindent Until recently, most PBH research focused on obtaining observational limits on their number density. The limits span an enormous mass range, from $10^{-5}\.$g to $10^{14}\.\Msun$, and derive from diverse astrophysical and cosmological phenomena. They encompass processes such as quantum evaporation, gravitational lensing, dynamical interactions, accretion, and gravitational-wave emission. As shown in Fig.~\ref{fig:constraints}, they can be expressed as upper limits on $\fPBH( M )$, the fraction of the dark matter in PBHs with mass $M$. The constraints are summarised briefly below and discussed in more detail in~\citet{2021RPPh...84k6902C}. 

\paragraph*{Evaporation}
\vs{2mm}

\begin{figure}[t]
	\centering
	\includegraphics[width=0.7\textwidth]{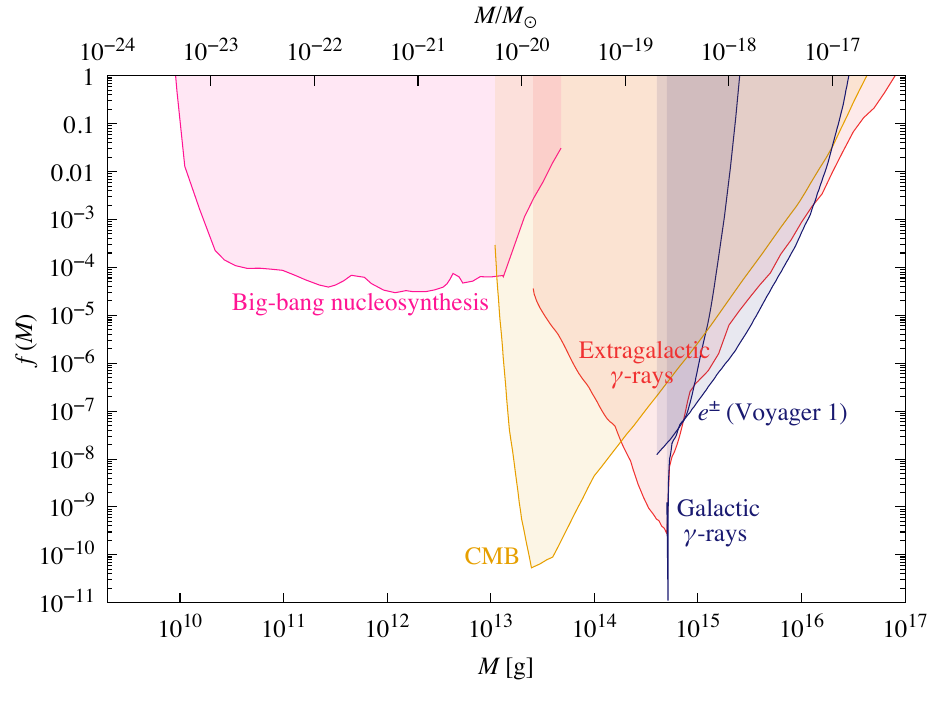}
	\caption{%
		Evaporation constraints on $\beta( M )$, 
		the fraction of Universe collapsing into 
		PBHs of mass $M$.
		From~\citet{2021RPPh...84k6902C}.}
	\label{fig:bc2}
\end{figure}

\noindent PBHs of mass $M$ will emit any elementary particles with rest mass below the temperature given by Eq.~\eqref{eq:temp}. For those lighter than $2 \times 10^{14}\.\grm$, this temperature exceeds the QCD confinement scale, resulting in the emission of quark and gluon jets that subsequently fragment into stable particles. The overall radiation therefore comprises both these primary emissions and their secondary products. \citet{MacGibbon:1991vc} first studied the observational consequences of this, in particular the possible PBH contribution to the extragalactic $\gamma$-ray background and cosmic rays. This led to stringent constraints on $\beta( M )$ and $\fPBH( M )$ for $M \lesssim 10^{17}\.\grm$, with PBHs of mass $M_{*} \approx 5 \times 10^{14}\.\grm$ completing their evaporation at the present epoch. These constraints were extended and improved by numerous later studies and their status in 2021 is shown in Fig.~\ref{fig:bc2}. This updates the BBN bound and shows the constraint from the extragalactic and Galactic $\gamma$-ray backgrounds, the spectrum and isotropy of the CMB and Voyager I measurements of the electron/positron cosmic rays. The public release in 2019 of the \texttt{BlackHawk} code, which calculates the evaporation spectra produced by any population of PBHs, led to improved limits on those lighter than $10^{17}\.\grm$~\citep{Auffinger:2022khh}.

\paragraph*{Lensing}
\label{subsec:lensing}
\vs{2mm}

\noindent Stellar microlensing provides a powerful tool for detecting planetary to solar-mass compact objects. This manifests as a temporary wavelength-independent brightening of background stars when compact objects traverse their line of sight~\citep{Paczynski:1985jf}. Major observational campaigns began in the 1990s with the EROS, MACHO and OGLE surveys of the Milky Way. The MACHO observations of stars in the Magellanic Clouds revealed an unexpectedly high number of lensing events and initially suggested that compact objects of $0.5\.\Msun$ might constitute much of the halo mass~\citep{MACHO:1996qam}. BBN constraints on the baryon density excluded such a large population of astrophysical objects (e.g.~white dwarfs), so PBHs were a plausible explanation. However, later MACHO results{\,---\,}under the assumption that the halo is an isothermal sphere with a flat rotation curve{\,---\,}reduced the best-fit halo fraction, with the absence of long-duration events excluding $1${\,--\,}$30\.\Msun$ compact objects from comprising the dark matter~\citep{MACHO:2000qbb}. The results from EROS and OGLE gave even stronger constraints and appeared to exclude planetary and stellar mass compact objects making up more than $10\%$ of the halo but we return to this claim later.

\paragraph*{Gravitational Waves}
\vs{2mm}

\noindent If PBHs constitute a major component of dark matter, they would naturally form binary systems in the early Universe. 
Nearby pairs can become gravitationally bound before matter-radiation equality, with neighbouring PBHs inducing orbital eccentricity through gravitational torques. These binary systems could persist until the present, and those with a merger timescale comparable to the age of the Universe would produce detectable gravitational-wave signals, provided they remain intact within galactic halos. Gravitational waves from merging black holes were detected for the first time in 2015 and there are now nearly a hundred such events. We discuss whether some of these could be PBHs later but, even if they result from stars, one can place limits on $\fPBH$ for a wide range of masses. The constraint shown in Fig.~\ref{fig:constraints} is taken from \citet{2021JPhG...48d3001G}.

\paragraph*{Dynamics}
\vs{2mm}

\noindent There are numerous constraints related to the dynamical effects of PBHs. These are associated with the heating of galactic disks, effects on dwarf galaxies, triggering of white dwarf explosions, the swallowing of PBHs by neutron stars, dynamical friction effects, tidal distortions in galaxies and the CMB dipole anisotropy. Many of them involve the heating or destruction of astronomical systems by the passage of nearby PBHs. For example, when compact objects interact with wide binary star systems, they transfer energy to them, increasing their separation and unbinding the widest ones. A comparison of theoretical predictions with the observed distributions of wide binaries excludes PBHs with $M \gtrsim 10\.\Msun$ from providing the dark matter in our own halo. These and other dynamical constraints are shown in Fig.~\ref{fig:constraints}.

\paragraph*{Accretion and Mixed Dark Matter}
\vs{2mm}

\noindent PBHs with $M \gtrsim 10\.\Msun$ can significantly impact their surroundings through accretion of surrounding gas. This could alter the Universe's recombination history and generate anisotropies and spectral distortions in the CMB~\citep{Ricotti:2007au}. The emission of X-rays from PBH accretion provides constraints over the mass range $10 - 10^7\.\Msun$. In mixed scenarios, where the dark matter consists of both PBHs and WIMPs, one expects an ultracompact minihalo of WIMPs to surround each PBH. The WIMPs would annihilate, producing $\gamma$-rays, and observations then imply that PBHs and WIMPs cannot both contribute substantially to the dark matter~\citep{Adamek:2019gns}. 
On the one hand, confirmation of PBH dark matter would exclude an appreciable density of WIMPs. On the other hand, confirmation of WIMP dark matter would exclude a sizeable amount of PBHs. Numerical simulations of particle dark matter distributions around PBHs 
has refined our understanding of these hybrid models.

\section*{Evidence for PBHs}
\label{sec:Evidence-for-PBHs}

\begin{figure}
	\vs{-3mm}
	\centering
	\includegraphics[width=0.5\textwidth]{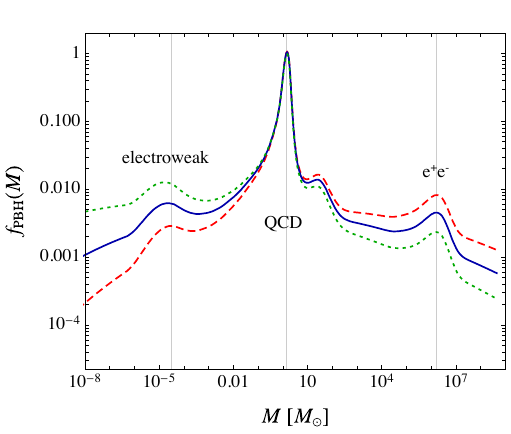}
	\caption{%
		The mass spectrum of PBHs for a spectral 
		index $n_{\srm} = 0.965$ (red dashed),
		$0.97$ (blue solid), $0.975$ (green 
		dotted) and running 
		$\alpha_{\srm} = - 0.0018$. 
		The grey vertical lines corresponds to 
		the EW and QCD phase transitions 
		and $e^{+}e^{-}$ annihilation. 
		From~\citet{Carr:2019kxo}. 
			} 
	\label{fig:fPBH}
\end{figure}

\begin{figure}[t]
	\centering
	\includegraphics[width=0.7\textwidth]{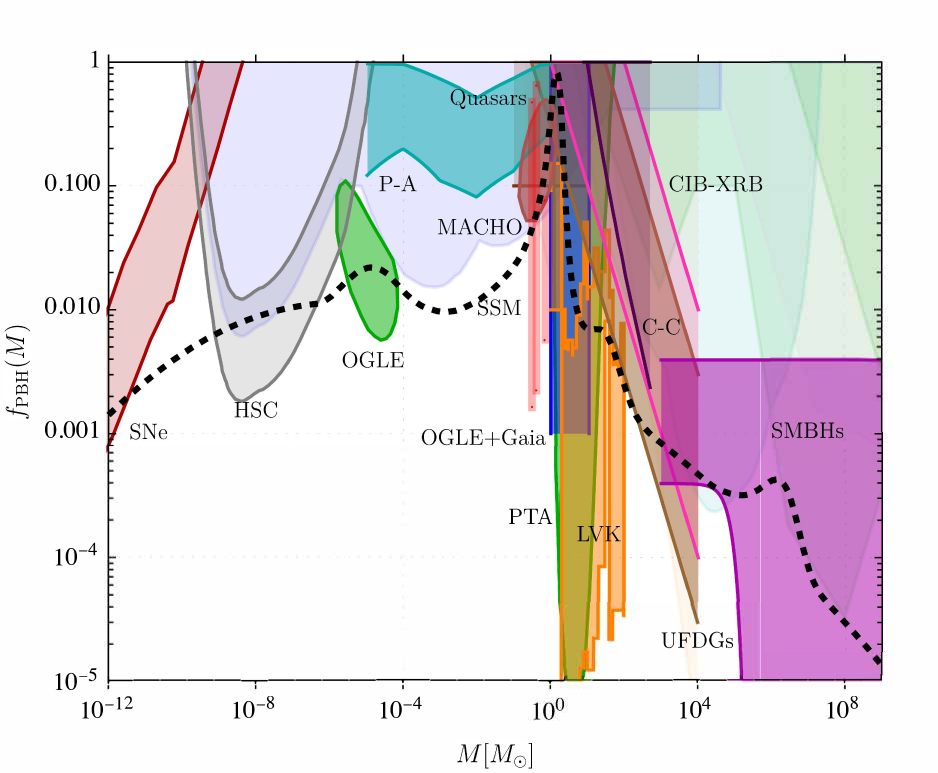}
	\caption{%
		Values of $\fPBH$ required by claimed 
		positive evidence. The evidence comes 
		from PBH-attributed signals from 
			supernov{\ae} (SNe), 
			various microlensing surveys 
			({\it Gaia}, HSC, OGLE, MACHO), 
			POINT-AGAPE pixel-lensing (P-A), 
			gravitational waves (LVK), 
			ultra-faint dwarf galaxies (UFDGs), 
			supermassive black holes (SMBHs), 
			core/cusp (C-C) profiles for inner 
			galactic halos, 
		and	
			correlations of the source-subtracted 
			cosmic infrared and X-ray backgrounds 
			(CIB-XRB).
		Also shown is the PBH mass function 
		induced by the thermal history of the 
		Universe (thick dashed curve) and the 
		constraints (lightly shaded). 
		From~\citet{Carr:2023tpt}.
		}
	\label{fig:figZ1}
\end{figure}

\noindent The discussion so far has focussed on PBH constraints but in this section we turn to some of the evidence which has accumulated in recent years. This corresponds to what is termed a ``positivist'' approach to the subject~\citep{Carr:2023tpt}. We will first discuss a very natural ``thermal history'' scenario for PBH formation at around the QCD epoch which gives the PBH mass function indicated in Fig.~\ref{fig:fPBH}. We then argue that much of the alleged evidence seems to support this, as illustrated in Fig.~\ref{fig:figZ1}. However, not all PBH advocates share our enthusiasm for this model and some would prefer a different PBH mass range.

\paragraph*{Thermal History Scenario}
\vs{2mm}

\noindent Reheating at the end of inflation fills the Universe with radiation. In the standard model, it remains dominated by relativistic particles and the number of relativistic degrees of freedom remains constant as time increases until around $200\.$GeV, when the temperature of the Universe falls to the mass thresholds of the Standard Model particles. The first particle to become non-relativistic is the top quark at $172\.$GeV, followed by the Higgs boson at $125\.$GeV, the $Z$ boson at $92\.$GeV and the $W$ boson at $81\.$GeV. At the QCD transition at around $200\.$MeV, protons, neutrons and pions condense out of the quarks and gluons. A little later the pions become non-relativistic and then the muons, with $e^{+}e^{-}$ annihilation and neutrino decoupling occurring at around $1\.$MeV.

Whenever the number of relativistic degrees of freedom suddenly drops, it changes the effective equation of state parameter $w$. There are thus four periods in the thermal history of the Universe when $w$ decreases for a short period. After each of these, $w$ resumes its relativistic value of $1 / 3$ but because the threshold $\delta_{\crm}$ is sensitive to the equation-of-state parameter $w( T )$, this modifies the probability of gravitational collapse of any large curvature fluctuations. This results in pronounced features in the PBH mass function even for a uniform power spectrum. If the PBHs form from Gaussian inhomogeneities, then Eq.~\eqref{eq:beta} implies that the fraction of horizon patches undergoing collapse when the temperature of the Universe is $T$ should be
\begin{align}
	\beta( M )
		\approx
			{\rm Erfc}\mspace{-1.5mu}
			\left[
				\frac{
					\delta_{\crm}
					\big(
						w[ T( M ) ]
					\big) }
				{ \sqrt{2}\,\epsilon( M )}
			\right]
			\. ,
			\label{eq:beta(T)}
\end{align}
where $\delta_{\crm}$ is the threshold for collapse, regarded as a function of the temperature, which is itself related to the PBH mass by $T \approx 200\,\sqrt{\Msun / M\,}$ $\MeV$. Thus $\beta( M )$ is exponentially sensitive to $w( M )$ and the present CDM fraction for PBHs of mass $M$ is 
\begin{align}
	\fPBH( M ) 
		\equiv
			\frac{ 1 }
			{ \rho_{\rm CDM} }
			\frac{ \drm\.\rho_{\rm PBH}( M ) }
			{ \drm \ln M }
		\approx
			2.4\;\beta( M )
			\sqrt{\frac{ M_{\rm eq} }{ M }\,}
			\, ,
			\label{eq:fPBH}
\end{align}
where $M_{\rm eq} = 2.8 \times 10^{17}\.\Msun$ is the horizon mass at matter-radiation equality and the numerical factor is $2\.( 1 + \Omega_{\Brm} / \Omega_{\rm CDM} )$. The resulting mass function is shown in Fig.~\ref{fig:fPBH}, using a power spectrum with three values for the spectral indices $n_{\srm}$ and a single value for the running $\alpha_{\srm}$. These values are chosen to be compatible with CMB measurements but with an amplitude such that the PBHs provide all the dark matter. It exhibits a dominant peak at $M \simeq 2\.\Msun$ and three additional bumps at $10^{-5}\.\Msun$, $30\.\Msun$ and $10^{6}\.\Msun$. 
 
\newpage

\paragraph*{Evaporations}
\vs{2mm}

\noindent In a series of papers~\citet{Cline:1996zg} have argued that exploding PBHs could explain some short-duration $\gamma$-ray bursts. Unlike cosmological $\gamma$-ray bursts, these would be located within the Galactic halo and therefore anisotropically distributed. This would require some new effect when the black hole temperature reaches the QCD scale. Contrary to earlier claims, evaporating black holes do not develop photospheres due to electron-positron or QCD interactions, 
so some extra effect would be required.

The anticipated clustering of evaporating PBHs within our halo should produce a distinctive Galactic $\gamma$-ray signature. The anisotropy of this Galactic background distinguishes it from the uniform extragalactic background. While the latter comes from the time-integrated emission of PBHs with initial mass $M_{*}$, the Galactic signal stems from the ongoing evaporation of slightly more massive PBHs that are now smaller than $M_{*}$ but have not evaporated completely. Specifically, the PBHs which generate the Galactic background have an initial mass $M_{*}( 1 + \mu )$ and a current mass $( 3 \mu )^{1/3} M_{*} $ with $\mu \ll 1$. \citet{1996ApJ...459..487W} once claimed that a Galactic background had been detected in EGRET observations between $ 30\.\mathrm{MeV} $ and $120\.\mathrm{MeV}$ and attributed this to PBHs. Later analysis of EGRET data, assuming a variety of PBH distributions, 
reassessed this limit by including a realistic model for the PBH mass spectrum and a more precise relationship between the initial and current PBH mass.

\paragraph*{Lensing}
\vs{2mm}

\noindent Although the MACHO collaboration detected $17$ microlensing events, which is larger than can be attributed to known stellar populations in the Milky Way or Magellanic Clouds, they concluded that this did not suffice to explain all the dark matter~\citep{MACHO:2000qbb}. However, this conclusion may be obviated if the PBHs are clustered and have an extended mass spectrum, both features being expected, and if the Galactic rotation curve is falling rather than flat. Indeed, positivists claim that the extended mass function predicted by the thermal history model, with a peak at around a few solar masses, could still provide all the dark matter. However, this is a controversial claim, with the EROS and OGLE groups coming to a different conclusion.

Using data from the five-year OGLE survey of microlensing events in the Galactic bulge, \citet{Niikura:2019kqi} have identified six ultra-short ones attributable to planetary-mass objects between $10^{-6}$ and $10^{-4}\.\Msun$. Although these could only contribute about $1\%$ of the dark matter, this is much more than expected for free-floating planets. \citet{Niikura:2017zjd} have also carried out an observation of Andromeda with the Subaru {\it Hyper Suprime-Camera} (HSC) and reported a single microlensing event by a compact body with mass in the range $10^{-11}\,\text{--}\,10^{-5}\.\Msun$. 

Perhaps the most convincing evidence for stellar-mass PBHs comes from microlensing events in the light curves of multiply-lensed quasars. The first detection of such an effect came from the quasar Q0957+561, which is split into two separate images by a massive galaxy along the line of sigh. Photometric monitoring showed that small brightness changes in one image were repeated in the second image a year later, which confirmed its identification as a gravitational lens. Although such lensing might be attributed to ordinary stars, in some cases the line of sight is too far away from the galaxy for this to be viable. This suggests that the observed microlensing of the quasar images is more likely due to a cosmological distribution of stellar-mass compact bodies making up the dark matter~\citep{Hawkins:2020zie}.

\paragraph*{Dynamical Effects \& Accretion}
\vs{2mm}

\noindent We have seen that numerous dynamical effects can be used to constrain PBHs. However, each such effect could potentially provide evidence for them. For example, \citet{1985ApJ...299..633L} have argued that $10^{6}\.\Msun$ black holes could explain the heating of stars in our Galactic disk, although this is now usually attributed to more conventional effects. Using $N$-body simulations to explore a halo model comprising both PBHs and particles, \citet{Boldrini:2019isx} have shown that cores can form as a result of dynamical heating of the cold dark matter through PBH infall and two-body processes, thereby resolving the cusp/core problem. 

The triggering of white dwarf explosions by PBHs can lead to observable signatures. In particular, it is exciting that some recently observed supernov{\ae}, the so-called ``calcium-rich transients'' do not trace the stellar density but are located off-centre from their host galaxies. Furthermore, they appear to originate from white dwarfs with masses of around $0.6\.\Msun$, well below the Chandrasekhar limit, and they occur predominantly in old systems. \citet{Smirnov:2022zip} argue that these transient events could have been triggered by collisions with PBHs with $10^{21}\.\grm < M < 10^{23}\.\grm$ and $10^{-3} < \fPBH < 0.1$. It is interesting that~\citet{Fuller:2017uyd} invoke the same mass range to explain how some $r$-process elements (i.e.~those generated by fast nuclear reactions) can be produced by the interaction of PBHs with neutron stars. This requires $\fPBH > 0.01$ in the mass range $10^{19}${\,--\,}$10^{25}\.$g. 
Collisions of neutron stars with PBHs of mass $10^{23}\.$g may also explain the millisecond durations and large luminosities of fast radio bursts. 

Compelling evidence for solar-mass PBHs may also come from the spatial coherence of the source-subtracted cosmic infrared and X-ray backgrounds~\citep{2013ApJ...769...68C}. The level of the infrared background suggests an overabundance of high-redshift halos that could be explained by the PBH Poisson effect if a significant fraction of the CDM comprises stellar-mass PBHs. In these halos, a few stars form and emit infrared radiation, while the PBHs emit X-rays due to accretion. PBHs naturally explain both the amplitude and angular spectrum of the source-subtracted infrared anisotropies.

\paragraph*{Gravitational Waves}
\vs{2mm}

\noindent The first direct detection~\citep{Abbott:2016blz} of gravitational waves, originating from the merging of two black holes, triggered an avalanche of interest in a possible primordial origin. Within a month it was claimed that the expected binary PBH merger rate was compatible with the data if PBHs provide the dark matter, 
although others argued that the PBHs required would have much less than the dark matter density. 
This depends on whether the binaries formed at early times or much later in the Galactic halo.

Currently almost a hundred merger detections have been made, some of which are in mass gaps which are hard to explain with stellar precursors. For example, several events would need to involve a progenitor star with mass above $\sim 60\.\Msun$. In this case, the temperature in the stellar core becomes so high during oxygen-burning that electron-positron pair production leads to instability and core collapse. Below a mass of around $150\.\Msun$, the core explodes as a remnantless supernova, so stars are not expected to form black holes in the {\it upper} mass gap between $60\.\Msun$ and $150\.\Msun$. Other events involve objects in the {\it lower} mass gap between $2\.\Msun$ (the highest mass neutron star) and $5\.\Msun$ (the lowest mass black hole). Black holes in this mass gap are not usually observed, and there may be theoretical reasons why this is not expected for conventional stellar models. On the other hand, PBHs can naturally form within both these mass ranges.

The LVK data includes several binaries with a possible subsolar secondary component, these having high signal-to-noise ratio ($> 8$) and low false-alarm-rate ($< 2$ per year) at the usual LVK detection thresholds. Recently, three additional subsolar triggers have been reported. A primordial origin of subsolar compact objects would undoubtedly be the most plausible interpretation, as all other proposed explanations are more exotic. Another clear indications of a primordial origin would be the detection of a black hole merger at a redshift before the epoch of first star formation (viz.~$z > 20$). For chirp masses above or around $30\.\Msun$, such redshifts will become accessible with the third generation of ground-based gravitational-wave detectors, such as the {\it Einstein Telescope} and {\it Cosmic Explorer}.

Additional evidence for PBHs is the non-zero effective spins in the most recent gravitational-wave catalogue. Most stellar models should produce a wide distribution in the effective spin $\chi_{\rm eff}$ with a peak between zero and one. On the other hand, low spin is a generic outcome for PBHs formed from inflationary overdensities. Although coherent accretion, previous mergers and hyperbolic encounters in dense PBH environments are expected to broaden the spin distribution, this should still result in a distribution of $\chi_{\rm eff}$ centred around zero for coalescences in the pair-instability mass gap.

The NANOGrav collaboration{\,---\,}using pulsar timing arrays{\,---\,}has recently reported the detection of a GW background at nano-Hertz frequencies. This has been confirmed by the {\it European Pulsar Timing Array} and {\it International Pulsar Timing Array} collaborations, who find the expected Hellings--Downs signature. Although this background is commonly attributed to the merging of SMBHs, other studies have suggested that the signal could be associated with PBHs in the stellar or planetary-mass range.

\paragraph*{Supermassive Black Holes and Cosmic Structures}
\vs{2mm}

\noindent The possible role of PBHs as seeds for early cosmic structures and the SMBHs in galactic nuclei has been highlighted by recent observations. The standard view in the CDM picture is that all cosmic structures arise from inflationary fluctuations and that the SMBHs in galactic nuclei form after their host galaxies as a result of dynamical and accretion processes. However, both these assumptions are challenged by the JWST data, gravitationally bound star-forming regions (``little red dots'') and enormous black holes apparently forming earlier than expected in the usual CDM picture. For example, there are galaxies with mass above $10^{10}\.\Msun$ and central SMBHs at redshifts in the range $7.4$\,--\,$9.1$ 
and the detection of X-rays from a $4 \times 10^{7}\.\Msun$ SMBH in a gravitationally-lensed galaxy at $z = 10.3$ has been reported. PBHs may help to resolve these problems on account of two closely related effects. 
First, the Poisson fluctuations in the PBH number density generate overdensities larger than the standard CDM ones on small scales; this produces bound dark clusters of PBHs, with gas later falling inside them and forming stars. Second, the Coulomb effect of an individual PBH inevitably binds ever-larger regions{\,---\,}seeding either the central SMBH or the galaxy itself. Indeed, the SMBH could itself seed the galaxy and this naturally explains the observed proportionality between the mass of the SMBH and the host galaxy. It also implies that this ratio will be larger at earlier times and there may be evidence for this. The presence of a heavy seeding channel for the formation of SMBHs within the first billion years of cosmic evolution has been suggested 
on the basis of JWST results.

\section*{Conclusions and Outlook}
\label{Last-Word:-The-Even-Brighter-Future}

\noindent The field of PBH research stands at a crucial juncture, with definite evidence for or against their existence likely emerging within the coming decade. While researchers differ in their predictions, the unprecedented surge in PBH research has driven increasingly sophisticated analyses of their theoretical properties and observational signatures. Among those advocating PBHs as dark matter candidates, a significant divide exists between proponents of asteroidal-mass and solar-mass PBHs. The asteroidal-mass hypothesis has the advantage that it is not excluded by current observational constraints. The solar-mass proposal, though supported by various observations, faces more significant theoretical challenges due to existing constraints in this mass range.


\bibliographystyle{Harvard}
\bibliography{reference}
\end{document}